\documentclass[%
 reprint,
floatfix,
amsmath,amssymb,
aps,
]{revtex4-2}

\usepackage{xcolor}
\usepackage{diagbox}

\newcommand{\avg}[1]{\left\langle#1\right\rangle} 

\usepackage{graphicx}
\usepackage{dcolumn}
\usepackage{bm}

\begin{document}

\preprint{APS/123-QED}

\title{Less is different: why sparse networks with inhibition differ from complete graphs}

\author{Gustavo Menesse$^{1,2,3}$}
\email{mereles930@gmail.com}
\thanks{Corresponding author}
\affiliation{$^2$Departamento de Electromagnetismo y Física de la Materia, Facultad de Ciencias, University of Granada, 18071 Granada, Spain}
\altaffiliation[Also at]{$^1$Departamento de Física, FFCLRP, Universidade de São Paulo, Ribeirão Preto, SP, 14040-901, Brazil}
\altaffiliation[Also at]{$^3$Departamento de Física, FACEN, Universidad Nacional de Asunción, San Lorenzo, Paraguay}

\author{Osame Kinouchi$^{1,}$}
\email{osame@ffclrp.usp}
\affiliation{$^1$
 Departamento de Física, FFCLRP, Universidade de São Paulo, Ribeirão Preto, SP, 14040-901, Brazil
}%

\date{\today}

\begin{abstract}

In neuronal systems, inhibition contributes to stabilizing dynamics and regulating pattern formation. 
Through developing mean field theories of neuronal models, using complete graph  networks, 
inhibition is commonly viewed as one ``control parameter'' of the system, promoting an absorbing phase
transition. Here, we show that for low connectivity sparse networks, inhibition weight is
not a control parameter of the transition. We present analytical and simulation results using generic stochastic integrate-and-fire neurons that, under specific restrictions, become other simpler stochastic neuron models common in literature, which allow us to show that our results are valid for those models as well. We also give a simple explanation about why the inhibition role depends on topology, even when the topology has a dimensionality greater than the critical one. The absorbing transition independence of the inhibitory weight may be an important feature of a sparse network, as it will allow the network to maintain a near-critical regime, self-tuning average excitation, but at the same time, have the freedom to adjust inhibitory weights for computation, learning, and memory, exploiting the benefits of criticality.


\end{abstract}

\maketitle

The absorbing phase transition (AT) is the most explored type of transition in the brain criticality hypothesis \cite{niebur2014}, which proposes that biological neuronal networks operate around a critical regime to optimize information processing and stimuli sensibility~\cite{Kinouchi2006}. To increase the biological plausibility of neuronal models, inhibitory coupling is needed, and in doing so, a rich dynamical behavior emerges even in simple models \cite{Corral_2022}. In the literature, the use of complete graph (CG) topology is common for deriving analytical results for excitatory/inhibitory neuronal models \cite{Benayoun2010,Decandia2021,Piuvezam2023}. However, there is evidence that topologies other than CG give different network dynamics when inhibition is present \cite{Larremore2014,Angulo-Garcia_2017,Buendia2019}. Even in a random sparse graph, when inhibition is considered, the richness of neuronal dynamics greatly increases, and some unexpected behaviors emerge. Some of these interesting phenomena are the ``ceaseless activity" \cite{Larremore2014}, the activity rebirth \cite{Angulo-Garcia_2017}, and the low-activity intermediate (LAI) phase \cite{Buendia2019}, all of them caused by the introduction of inhibition into a system with a sparse topology (low connectivity).

In a now classical article \cite{Brunel2000}, Brunel explores the dynamics of sparse excitatory/inhibitory neuronal networks presenting a classification for neuronal dynamical regimes. In this seminal work, a mean-field-like theory was developed for a sparse network, and it was shown that different oscillatory regimes emerge controlled, among others, by relative synaptic inhibitory currents $g$. One of the regimes described by Brunel, the Asynchronous Irregular (AI) dynamics, is the center of a discussion about whether the cortex neuronal activity is critical or AI. Shew \textit{et al.} \cite{Shew2020} show in a similar model used by Larremore \textit{et al.} \cite{Larremore2014} and Buendia \textit{et al.}  \cite{Buendia2019} that increasing $g$ could turn a critical regime into an AI regime. This and other results will be discussed in this article through the lens of a simple but generic stochastic neuron model.

In recent articles, such as \cite{Bi_2021,diVolo_2022}, an extended discussion of the dynamics of inhibitory/excitatory neuronal sparse networks was presented. Mean-field theories of high and low connectivity were developed, with homogeneous and heterogeneous coupling distributions, showing how the interaction between topology and excitatory/inhibitory dynamics yields a rich dynamical repertoire. These papers do not focus on the absorbent or silent phase and do not discuss the behavior of the absorbing phase transition in sparse networks, which is relevant in the context of the brain criticality hypothesis. Here, using a far more straightforward mathematical approach, we will develop a tree-like mean-field theory to study AT in a sparse neuronal network model when both excitatory and inhibitory neurons are present. 

We consider here a network of $N$ discrete-time stochastic leaky integrate-and-fire neurons~\citep{Gerstner1992,Galves2013,Larremore2014,Costa2017,Zierenberg2020} considering both excitatory and inhibitory neurons ~\citep{Girardi-Schappo2021}. A Boolean indicator $X^{E/I}_i \in \{0,1\}$, $i = 1,\ldots,N_E$ or $N_I$,
denotes silence ($X^{E/I}_i = 0$) or the firing of an action potential (spike, $X^{E/I}_i = 1$), where the superscripts $E/I$ indicate the excitatory/inhibitory nature of the neuron.

When a neuron $i$ is inactive ($X^{E/I}=0$), the membrane potential evolves according to:
\begin{eqnarray}\label{eq:Pot}
    V^{E/I}_{i}[t + 1] &=& \Bigg[ \mu_i V^{E/I}_{i}[t] + I_i  \nonumber \\ &+& \left. \frac{1}{K}\left(
    \sum_{j=1}^{K_E}  J_{ij} X^{E}_j[t] -
    \sum_{j=1}^{K_I}  W_{ij} X^{I}_j[t] \right) \right] \nonumber \\ & & \cdot \left[1 - {X_{i}}^{E/I}[t]\right],
\end{eqnarray}
where $0 \leq \mu_i \leq 1$ are leakage parameters and 
$I_i$ are external inputs. We use the $[t]$ notation for
discrete-time.
Each neuron has $K_E$ excitatory and $K_I$ inhibitory neighbors, totaling $K=K_E+K_I$ incoming links. Outgoing links, by this
construction, have a binomial distribution with an average $K$
and a standard deviation
$\sigma  = \sqrt{ K(1-K/(N-1))}$.

If in time step $t$ the neuron fires, its voltage is reset $V^{E/I}_i[t + 1] = 0$. The transition between states from state $X[t]$ to $X[t+1]$, in general, will depend on the voltage $V_{i}$ following a transition probability matrix (Table \ref{Tab:1}),

\begin{table}[ht]\centering\setlength\tabcolsep{3.5pt}\renewcommand\arraystretch{1.25}
\caption{Transition probability matrix of the neuron states}
    \begin{tabular}{c|c|c}
      \diagbox[width=\dimexpr \textwidth/8+2\tabcolsep\relax, height=1cm]{ t }{t+1}
                   & $X^{E/I}=0$ & $X^{E/I}=1$ \\
      \hline
      $X^{E/I}=0$ &  $1-\Phi(V^{E/I}[t])$ & $\Phi(V^{E/I}[t])$ \\
      \hline
      $X^{E/I}=1$ & $\varphi(V^{E/I}[t])$ & $1-\varphi(V^{E/I}[t])$ \\
    \end{tabular}
 \label{Tab:1}
\end{table}

This means, a firing occurs with probability:
\begin{eqnarray}
    \label{eq:probfire}
    \mathbb{P}\left( X^{E/I}_i[t+1] = 1 \:|\: X^{E/I}_i[t] = 0, V^{E/I}_i[t]\right) \equiv \nonumber \\  \Phi_i(V^{E/I}_i[t])\:,
\end{eqnarray}
where $\Phi(V)$ is the so-called firing function. When a neuron fires, it will return to the inactive state (inactivation probability) with a probability
\begin{eqnarray}
    \label{eq:probdeactivate}
    \mathbb{P}\left( X^{E/I}_i[t+1] = 0 \:|\: X^{E/I}_i[t] = 1, V^{E/I}_i[t]\right) \equiv \nonumber \\  \varphi_i(V^{E/I}_i[t])\:.
\end{eqnarray}
The probability of staying inactive (not firing) and staying active (double firing) is the complement of the firing probability and the inactivation probability, respectively.

This model will incorporate
an absolute refractory period of one-time step after a spike if we impose
$\varphi(V^{E/I}[t])= 1$ and $\Phi(0) = 0$. This condition makes the probability of double firing equal 0, so the generic model becomes the Gerstner-Galves-Löcherbach (GGL) model~\citep{Gerstner1992,Galves2013,Brochini2016,Costa2017}.

Otherwise, if we impose the probability of inactivation $\varphi(V^{E/I}[t]) = 1 - \Phi_i(V^{E/I}[t])$, the normalization requirement makes the probability of double firing equal to $\Phi_i(V^{E/I}_i[t])$ and we will not have a refractory period so that the generic model becomes the Larremore \textit{et al.} model \cite{Larremore2014}.

Finally, we can also obtain a model with $n$ time-step refractory period if we use a firing probability with a refractory period control factor like $\tilde{\Phi}(V^{E/I}[t],t_{sp}) = \Phi(V^{E/I})\Theta( t- t_{sp}-n ) $ where $t_{sp}$ is the time of the last spike. 

\begin{figure}[ht]
    \centering
    \includegraphics[width=0.8\linewidth]{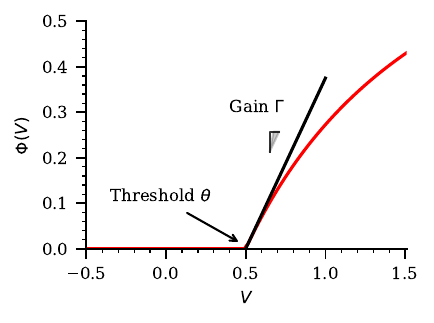}
    \caption{Firing function $\Phi(V)$. In red the 
    Eq.~\eqref{eq:phi}, in black is shown the inclination in the threshold point $\theta$ controlled by the gain $\Gamma$.}
    \label{fig:1}
\end{figure}

As for the GGL model, for the generic model, there are no strong requirements on the firing function $\Phi$ besides a sigmoid shape \citep{Brochini2016} and a firing threshold, the minimum voltage value needs to have a non-zero firing probability. One example of this is the so-called rational firing function,
see Fig.~\ref{fig:1}:
\begin{equation}
\label{eq:phi0}
\Phi_{i}(V^{E/I}_{i})= \frac{\Gamma_{i}\left( V^{E/I}_{i}-\theta_{i}  \right)}{1+\Gamma_{i}
\left( V^{E/I}_{i}-\theta_{i}  \right)} \:\:\Theta\left(V^{E/I}_{i}-\theta_{i}\right)\:,\end{equation}
where $\theta_i$ is the firing threshold, $\Gamma_i$ is the gain and $\Theta(x)$ is the Heaviside step function. Another commonly used function in the literature is the Linear Saturating Function \cite{Larremore2014,Girardi-Schappo2021,Buendia2019}. In general, the firing function has a form like,
\begin{equation}
\label{eq:phi}
\Phi_{i}(V^{E/I}_{i})= f\left( V^{E/I}_{i}-\theta_{i}  \right)\:\:\Theta\left(V^{E/I}_{i}-\theta_{i}\right)\:,\end{equation}
where $f$ is a continuous increasing function that tends to $0$ in the limit of $ (V^{E/I}_{i}-\theta_{i}) \rightarrow 0^{+}$ and to $1$ when $ (V^{E/I}_{i}-\theta_{i}) \gg 1$.

The order parameter of the systems is the time average of the fraction of spiking neurons (firing density) $\rho[t] = \left\langle X_i[t]\right\rangle \equiv \frac{1}{N}\sum_{i=1}^N X_i[t]$. In simulations, its time average $\displaystyle \rho^{*} = \avg{\rho[t]}_t $ is calculated after disregarding transients.

As shown in \cite{Girardi-Schappo2021}, it is possible to derive a mean-field  (MF) calculation that is exact for complete graph networks with self-averaging parameters. This calculation predicts
an AT. The leakage parameter $\mu$ does not change the properties of the phase transition such as the critical exponents, universality class, etc. but only the location of the transition point~\cite{Menesse2021}. So, we examine the case $\mu = 0$, where MF leads to the exact self-consistent equation (full analytical derivation of the CGMF self-consistent equation is presented in Supplemental Material I.A):
\begin{eqnarray}\label{eq:selfcons_CG}
\rho &=& -2\Gamma\bar{W}\rho^{2} + \left(\Gamma\bar{W}-2\Gamma h \right)\rho + \Gamma h \:,
\end{eqnarray}
where $\bar{W} = pJ-qW$, $h = I-\theta$, $p=N_E/N$ and $q=N_I/N = 1-p$ 
are the fractions of excitatory and inhibitory populations. Here, we define $ J = \avg{J_{ij}}$, $W=\left\langle W_{ij} \right\rangle$, 
$I=\avg{I_i }$, $\theta= \avg{\theta_i}$ as the average over neurons.

Solving Eq.~\eqref{eq:selfcons_CG} yields the stationary activity \cite{Girardi-Schappo2021}:
\begin{eqnarray}\label{eq:rho_CG}
\rho^{*} &=& \frac{\Gamma\bar{W}- 2\Gamma h -1}{4\Gamma\bar{W}} \nonumber \\  &  &\pm \frac{\sqrt{\left[\Gamma\bar{W}-2\Gamma h - 1\right]^{2}+8\Gamma^{2}\bar{W}h}}{4\Gamma\bar{W}} \: ,
\end{eqnarray}

An active ($\rho>0$) to absorbing ($\rho=0$) phase transition occurs in the limit of zero effective external field $h\rightarrow 0$. In this case, Eq.~\eqref{eq:rho_CG} yields to an absorbing (``silent'') phase ($\rho^{*}=0$) and an active phase
\begin{equation}
    \rho^{*} = \frac{1}{2}\frac{\left(\bar{W} - \displaystyle\frac{1}{\Gamma}\right)}{\bar{W}} \sim \left( \bar{W} - \bar{W}_c \right)^{\beta} \: .
\end{equation}

Previous work shows that the critical exponent $\beta = 1$ obtained from the mean-field approximation belongs to the directed percolation mean-field universality class \cite{Girardi-Schappo2021}. The critical surface $\displaystyle \bar{W_c} = 1/\Gamma$ can be expressed in terms of the synaptic coupling ratio $\displaystyle g=W/J$ \cite{Girardi-Schappo2021}, which gives: 
\begin{equation}\label{eq:Crict_GC}
    g_c = \frac{p}{q}-\frac{1}{q\Gamma J} \: .
\end{equation}
The use of the control parameter $g$ is usual
in the literature on balanced networks~\cite{Brunel2000}.
\begin{figure*}[htb]
    \centering
    \includegraphics[width=.325\linewidth]{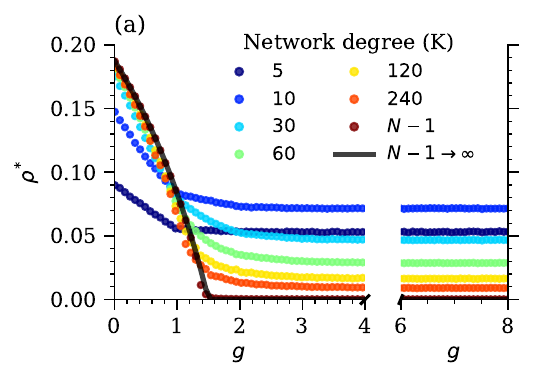}
    \includegraphics[width=.325\linewidth]{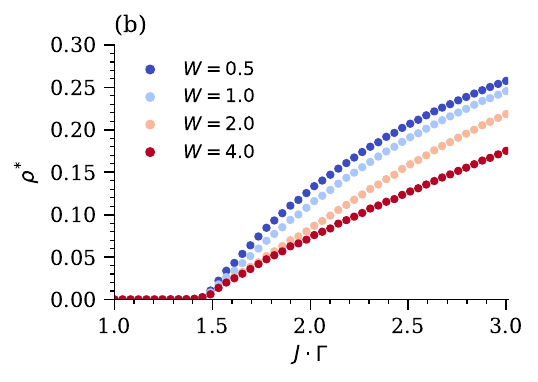}
    \includegraphics[width=.325\linewidth]{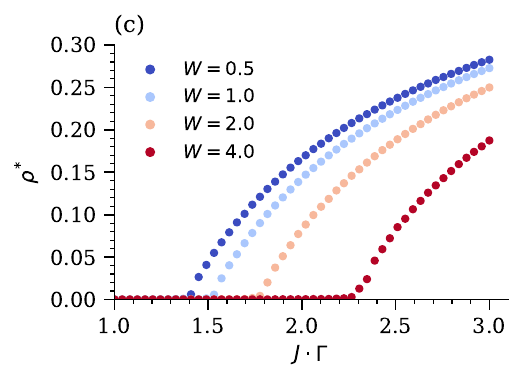}
    \caption{Steady state activity $\rho^{*}$ as a function of: (a) synaptic coupling ratio $g$ and (b,c) product between excitatory weights and gain. Simulations of the model in (a,b) random $K$-regular network and (a,c) complete graphs ($K=N-1$) with $N=10000$, $\Gamma=1$, $I=\theta=0$ and $\mu=0$. Complete graph MF \cite{Girardi-Schappo2021} (solid line) and simulations (dots). (a) The complete graph MF does not accurately describe the behavior of sparse networks ($K\ll N$), the activity becomes independent of $g$ for $g>1$ when $K$ is small. (b) In random sparse networks, the intensity of activity is modulated by $W$ in the active phase, but the absorbing transition line does not depend on it. (c) In the complete graph, $W$ affects the critical point value, so it is a control parameter given a fixed value of $J$ and $\Gamma$.}
    \label{fig:2}
\end{figure*}
This result indicates that, given $p$, $q$, $\Gamma$, and $J$ values, there is an inhibitory strength $W_c$ where the AT takes place. This leads us to think that inhibition is a control parameter of the system, an assumption that seems intuitive and general. However, we will show here that it is only valid in the limit $K\rightarrow N-1$ (complete graph).

The absorbing transition of this model is common in the brain criticality literature \cite{Gros_2021}. In excitatory networks, different topologies (small world, random graphs, etc.) were explored, and all of them agreed with the simple complete graph mean field analytical results \cite{Costa2017}. But when inhibition is added to the networks, the literature on the GGL model typically only shows agreement with complete graph simulations \cite{Carvalho2021,Girardi-Schappo2021}.

Using a directed $K$-regular topology, we show that the complete graph MF result presented in Eq.~\eqref{eq:Crict_GC} does not agree with the simulations. As shown in Fig.~\ref{fig:2}a, for sparse networks, increasing the inhibitory weight does not promote the phase transition. Increasing the network in degree $K$, we see how the simulation results slowly converge to the complete graph mean field as $K\rightarrow N-1$, but the absorbing phase only emerges when the neurons of the network reach all-to-all interaction.

In Fig.~\ref{fig:2}b we can see that the inhibitory weights $W$ modulate the network activity only in the active phase, but do not have any influence on the location of the critical point. This must be compared with
CG networks
(Fig.~\ref{fig:2}c) where the same parameter $W$ drastically affects
the critical point.

To understand this behavior, we use a tree-like mean field approximation \cite{Bethe1935}. The important presumption for this mean field is that transition  probabilities  are  translationally  invariant  in  the thermodynamic limit and beyond the upper critical dimension, so we can use a representative arbitrary neuron and its $K_E$ and $K_I$ neighbors to describe the mean behavior of the network. Considering that in sparse networks the probability of having loops in the neighborhood of a node is low, we can assume that the network is locally a tree and thus, the activity of the neighbors of a neuron is statistically independent. These same intuitions were used before to study sparse neuronal networks as in ~\cite{Brunel2000,Buendia2019}.

From Eq.~\eqref{eq:Pot}, when $\mu=0$ and $h=0$, we estimate the stationary value $V^{*}$ of an inactive random neuron $i$ as:
\begin{equation}\label{eq:PotStat}
V^{*} = \frac{1}{K}\left(J\sum_{j}^{K_E}X_{j}^{E}-W\sum_{j}^{K_I}X_{j}^{I}\right) \:.
\end{equation}
Defining the number of active excitatory and inhibitory neighbors as $\sum_{j}^{K_E}{X_j}^{E}=m_E$ and $\sum_{j}^{K_I}{X_j}^{I}=m_I$, $\gamma = \Gamma J/K$ and $g=W/J$, the firing function (generic) of a random inactive neuron is:
\begin{equation}
\Phi(m_E,m_I) = \displaystyle f\left(\gamma\left( m_E - g m_I \right)\right)\Theta\left( m_E - g m_I \right) \:.
\end{equation}

Now, the independence between neighbors states allows us to express the probability of finding a combination of $\{m_E,m_I\}$ active neighbors as the product of binomial probabilities,
\begin{eqnarray}\label{eq:comb}
\mathbb{P}\left(\{m_E,m_I\}\right) &=& \binom{K_I}{m_I}\rho^{m_I}\left(1-\rho\right)^{K_I-m_I} \nonumber \\ &&\times  \binom{K_E}{m_E}\rho^{m_E}\left(1-\rho\right)^{K_E-m_E} \:.
\end{eqnarray}
Here, we use the supposition that the probability of finding an active neighbor is equal to the frequency of active neurons in the network, which is the same as the network activity $\rho$ defined before.

The mean value of the state of a neuron in the network at time $t$ is
\begin{eqnarray}\label{eq:Avg_rho}
\avg{X[t]}&=&\sum_{x\in\{0,1\}}x\mathbb{P}(X[t]=x_i)=\mathbb{P}(X[t]=1) \\
&\approx&\rho[t] \nonumber \:.  
\end{eqnarray}

The probability $\mathbb{P}(X[t+1]=1)$ of having an arbitrary neuron active in time $t+1$ has two contributions, one from the jump $0 \rightarrow 1$ (inactive at time t to active at time $t+1$) and the other from the probability of staying active $1 \rightarrow 1$, given all possible combinations of neighbor states $\{m_E,m_I\}$. Using the transition probabilities (Table \ref{Tab:1}), the Eq.~\eqref{eq:Avg_rho} and the fact that neuron states are independent at the same time $t$(causal locality), the activity dynamics can be described by,
\begin{align}\label{eq:rho_dyn}
\rho[t+1] &= \sum_{\{m_E,m_I\}}\left[(1-\rho[t])\Phi(\{m_E,m_I\}) \right. \nonumber \\ & +  \left. \rho[t]\left(1-\varphi(\{ m_E,m_I \})\right) \right]\cdot\mathbb{P}(\{m_E,m_I\}) \:,
\end{align}
a complete derivation of Eq.\eqref{eq:rho_dyn} is presented in Supplemental Material I.B.

The Eq.~\eqref{eq:rho_dyn} allows us to express the self-consistent equation ($\rho=\rho[t+1]=\rho[t]$) as:
\begin{eqnarray}
\label{eq:selfcons_SN}
\rho &=& \sum_{\{m_E,m_I\}}\left[(1-\rho)\Phi(\{m_E,m_I\}) \right. \nonumber \\ & &+  \left. \rho\left(1-\varphi(\{ m_E,m_I \})\right) \right]\mathbb{P}(\{m_E,m_I\})\:.
\end{eqnarray}

By one hand, if we impose $\varphi(\{ m_E,m_I \})=1$ and $\Phi(0) = 0$ the model incorporates an absolute refractory period of one-time step and became the GGL model,
\begin{eqnarray}
\label{eq:selfcons_SN_0}
\rho &=& \sum_{\{m_E,m_I\}}(1-\rho)\Phi(\{m_E,m_I\})\mathbb{P}(\{m_E,m_I\})\:.
\end{eqnarray}
By other hand, if we impose $\varphi(\{ m_E,m_I \}) = 1-\Phi(\{ m_E,m_I \})$ the model becomes the Larremore \textit{et al.} model and the self-consistency equation will be,
\begin{eqnarray}
\label{eq:selfconsLarre_SN_0}
\rho &=& \sum_{\{m_E,m_I\}}\Phi(\{m_E,m_I\})\mathbb{P}(\{m_E,m_I\})\:.
\end{eqnarray}
The only difference is the factor $1-\rho$ related to the existence of a one-time step absolute refractory period in the GGL model.

\begin{figure*}[htb]
    \centering
    \includegraphics{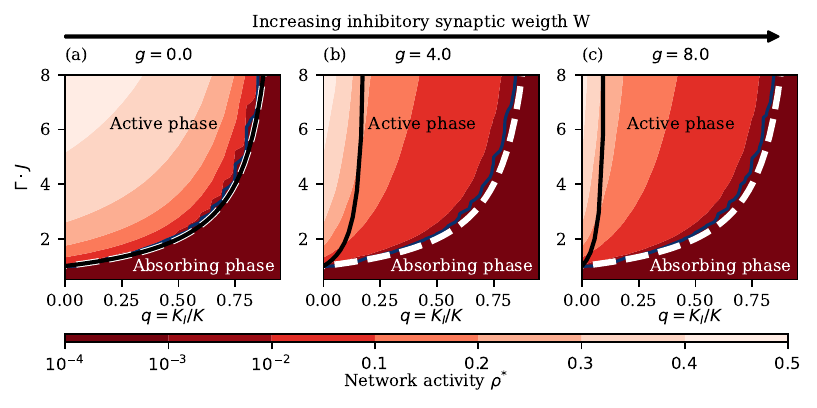}
    \caption{Phase diagram. Simulations in random $K$-regular network with $K=20$, $N=10000$, and different relative inhibitory weights $g=W/J$ with fixed $J=1$. The heat map shows the network stationary activity $\rho^{*}$, the dashed lines are the critical curve obtained analytically by tree-like MF approximation, and the solid black lines are the critical curves obtained by complete graph MF calculation. The absorbing transition is controlled by excitatory weight $J$, gain $\Gamma$, and the proportion of inhibitory neurons $q=K_I/K$, but not by the relative intensity of inhibitory weights $g=W/J$ as predicted by the complete graph MF. The blue line is the contour curve of $\rho^{*}=5\times10^{-4}$. Both mean-field approximations agree when there is no inhibition $g=0$.}
    \label{fig:3}
\end{figure*}

To obtain a meaningful analytical result from Eq.\eqref{eq:selfcons_SN_0}, having defined all factors, we can specify values of $K_E$ and $K_I$, then expand Eq.~\eqref{eq:selfcons_SN_0} to the second order to explore the transition region ($\rho\rightarrow 0^{+}$). Doing this for different values of $K_E$ and $K_I$ it is possible to infer the general result of the second-order expansion. However, we will do some more approximations to derive an analytical critical curve for the GGL model \eqref{eq:selfcons_SN_0}. 

First, we can rewrite the sum relative to active excitatory neighbors exploiting the Heaviside function, then
re-indexing the excitatory sum using $m^{\prime}_E=m_E-\lceil gm_I\rceil$ and $K^{\prime}_E=K_E-\lceil gm_I \rceil$, and approximating $\lceil gm_I \rceil \approx gm_I$. Following these steps, we obtain:
\begin{eqnarray}\label{eq:selfcons_SN_1}
\rho &=& (1-\rho)\sum_{m_I=0}^{K_I}\binom{K_I}{m_I}\rho^{m_I}\left(1-\rho\right)^{K_I-m_I} \\ &&\times \sum_{m^{\prime}_E=0}^{K^{\prime}_E}\binom{K^{\prime}_E}{m^{\prime}_E}\rho^{m^{\prime}_E+gm_I}\left(1-\rho\right)^{K^{\prime}_E-m_E} f(\gamma m^{\prime}_E)\:, \nonumber
\end{eqnarray}
were $m^{\prime}_E$ can be viewed as being the active effective excitatory neighbors (AEEN). To simplify the expression, we define the probability of firing under the influence of one AEEN as $\eta = f(\gamma)$, and $\bar{\eta}=1-\eta$ will be the probability of staying inactive under the same influence. Then, considering that the firings caused by different neighbors as mutually exclusive events, which is not true, but is valid for low activity. We can express the probability of firing when having $m^{\prime}_E$ AEEN as $1-\bar{\eta}^{m^{\prime}_E}$, so the firing function factor $f(\gamma m^{\prime}_E)$ will be reduced to $1-\bar{\eta}^{m^{\prime}_E}$. Using this, we can simplify the excitatory sum of Eq.~\eqref{eq:selfcons_SN_1} as follows:
\begin{widetext}
\begin{eqnarray}\label{eq:sim_E}
\sum_{m^{\prime}_E=0}^{K^{\prime}_E}\binom{K^{\prime}_E}{m^{\prime}_E}\rho^{m^{\prime}_E+gm_I}\left(1-\rho\right)^{K^{\prime}_E-m^{\prime}_E}\left(1-\bar{\eta}^{m^{\prime}_E}\right) &=& \rho^{gm_I}\left[1- \sum_{m^{\prime}_E=0}^{K^{\prime}_E}\binom{K^{\prime}_E}{m^{\prime}_E}\left(\rho\bar{\eta}\right)^{m^{\prime}_E}\left(1-\rho\right)^{K^{\prime}_E-m^{\prime}_E}\right] \nonumber \\
&=& \rho^{gm_I}\left[1-\left( 1-\eta\rho\right)^{K^{\prime}_E} \right]=\rho^{gm_I}\left[1-\left( 1-\eta\rho\right)^{K_E-gm_I} \right] \:.
\end{eqnarray}
\end{widetext}

Substituting Eq.~\eqref{eq:sim_E} in Eq.~\eqref{eq:selfcons_SN_1},
doing some algebra, writing $1-a=\rho^{g}$ and $1-b=\left(\rho/(1-\eta\rho)\right)^{g}$, we obtain a simplified self-consistency equation
\begin{equation}\label{eq:selfcons_SN_sim_02}
\rho = (1-\rho)\left[(1-\rho a)^{K_I}-\left(1-\eta\rho\right)^{K_E}\left(1-\rho b\right)^{K_I} \right]\:.
\end{equation}
The reader must notice that the Larremore \textit{et al.} model will have almost the same expression, with the only difference that the first factor $(1-\rho)$ will not be present. 

When $g>1$, substituting the original expressions for $a$ and $b$ and expanding \eqref{eq:selfcons_SN_sim_02} to second order about $\rho=0$ yields
\begin{equation}\label{eq:selfcons_SN_sim_3}
\rho^{2}\left(K_E\eta +\frac{K_E\eta^{2}\left(K_E-1\right)}{2}+K_I K_E\eta\right)+\rho\left(1-K_E\eta\right) = 0 \:,
\end{equation}
The terms that depend on $g$ disappear,
signalizing that the phase transition does not
depend on $g$. Solving Eq.~\eqref{eq:selfcons_SN_sim_3} we find the absorbing phase $\rho^{*}=0$, and
\begin{equation}
    \rho^{*}\approx G(\eta,K_E,K_I)\left(K_E\eta-1\right) \:,
\end{equation}
where $G(\eta,K_E,K_I)=\frac{2}{\eta K_E}\left[\frac{1}{2+K_I+\eta(K_E-1)}\right]$. 

Substituting $\eta$ for the original firing function factor $f(\gamma)$,

\begin{equation}
    \rho^{*}\approx G(\gamma,K_E,K_I)\left(K_E f(\gamma)-1\right) \:.
\end{equation}

If the rational firing function were used, $f(\gamma)=\frac{\gamma}{1+\gamma}$, where $\gamma=\frac{\Gamma J}{K}$. Then we find that
\begin{equation}
\rho^{*} \propto \left(\frac{\Gamma J - \frac{K}{K_E-1}}{\frac{K}{K_E-1}}\right)\sim \left(\frac{\Gamma J - (\Gamma J)_c}{(\Gamma J)_c} \right)^{\beta}\:,
\end{equation}
with $\beta = 1$.
The critical line is $\Gamma J=K/(K_E-1)= 
K/((K-1)-K_I)$. The line is independent of $g$, therefore, independent of synaptic inhibitory weights, as seen in the simulations. When $g<1$, the same critical line appears, but the function $G$ depends on $g$. The dependence on $g$ for $g<1$ and the independence for $g>1$ is exactly what we found in the simulations with $K\ll N$ (Fig.\ref{fig:2}.a, for $K=5$). 

When the linear saturating firing function is used, in the linear part of the function, we have $f(\gamma) = \gamma = \frac{\Gamma J}{K}$. This yields a critical line $\Gamma J = \frac{K}{K_E}$. An interesting fact here is that when there are no inhibitory neurons ($K=K_E$) the critical line is $\Gamma J = 1$. A peculiar regime in which AT is independent of local topology (critical curve independent of degree K), as shown in \cite{Menesse2021}.

Although inhibitory weight is not important for AT, we can see that the proportion of excitatory and inhibitory neurons plays an important role in AT. This is consistent with the literature, where connectivity between excitatory and inhibitory populations is considered a relevant control parameter to regulate the dynamic transitions of the neuronal network ~\cite{Poil2012,Avramiea2221}. 

In Fig. \ref{fig:3} we explore the phase diagram using the control parameter $\Gamma\cdot J$ and the local proportion of inhibitory neurons $\displaystyle q=K_I/K$. We can see how the derived critical line (dashed white curve) accurately describes the AT, while the CGMF critical line (solid white curve) does not. Simulation results presented by Buendia \textit{et. al} \cite{Buendia2019} show that the CGMF theory begins to agree with the system behavior when the ratio $K/N > 0.03$, which is not incompatible with what we observe here, we can use this as a criterion to differentiate low connectivity from high connectivity networks. 

\begin{figure*}[ht]
    \centering
    \includegraphics{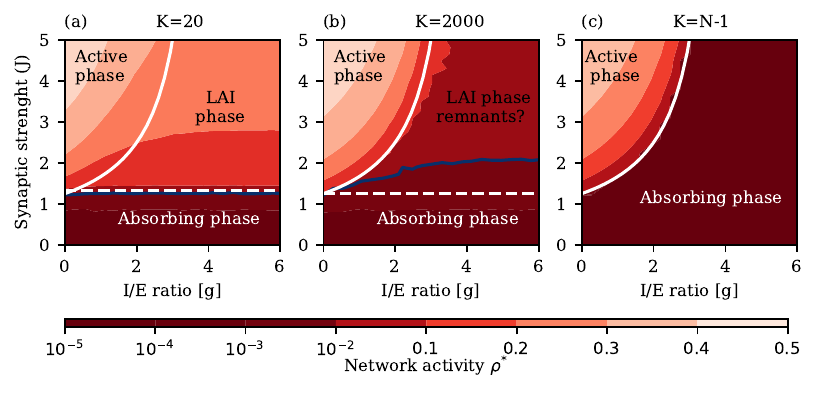}
    \caption{Phase diagram of the GGL model ($J$vs$g$) from low to the high connected sparse network until reaching a complete graph: Stationary activity obtained in simulations on $T=10000$ time-steps on a K-regular random network with $N=10000$, $I=\theta=0$, $\Gamma=1$ and inhibitory fraction $q=K_I/K=0.2$, (a) Low connectivity $K/N<0.03$ (b) High connectivity $K/N>0.1$ (c) Complete graph $K/N \approx 1$. White dashed line is the AT critical curve $\Gamma J_c=\frac{K}{K_E-1}$, the white solid is the CGMF critical curve $J_c=\frac{1}{(1-q)-qg}$, blue curve is the level curve of $\rho^{*}=10^{-3}$. Between the blue and green curves, we observed remnants of the LAI phase. The critical line for AT does not correctly describe the transition as the system has a high connectivity. The inhibitory current begins to control the AT as can be seen in the inclination of the blue line in (b). (c) In the complete graph, the LAI phase disappears and becomes part of the absorbing phase, and the AT transition is described by the CGMF.}
    \label{fig:4}
\end{figure*}

We can explore what happens with the phase diagram in terms of the control parameter $\Gamma J$ and the relative inhibitory current $g$ for different increasing values of $K$ (Fig.\ref{fig:4}). We can clearly see the AT independence of $g$ in a low connectivity sparse network, but this independence begins to break for high connected network until we reach the complete graph.

In the classical Brunel article \cite{Brunel2000}, some analytical approximations required a large degree $K$ (on the order of $10^3$), and simulations were done with networks of size $N=10000$, which combined with large $K$ considerably reduces the sparsity of the network, making it a highly connected network. However, in the same publication of Brunel and also in another publication of the same year\cite{BRUNEL2000b}, the author shows a phase diagram of a sparse neuron, where $K_E/N \ll 1$ is guaranteed. In this phase diagram, the bifurcation line where the almost quiescent state (equivalent to our absorbing transition) loses stability is also independent of $g$, in agreement with what happens in our model. In some sense, our results are already present in the Brunel model, but no discussion about it was done there, as the author focuses on the different oscillatory regimes that emerge in the active phase, which are controlled by $g$, but not in the effects of $g$ and network topology over AT, which is the main contribution of our present article.

In the Larremore \textit{et al.} model case, a similar Bethe-Peierls mean field approximation was developed in \cite{Buendia2019}, however, in that case the authors used a very specific firing function, a linear firing function with gain $\Gamma=1$. Furthermore, in that article, the authors do not separate the inhibitory from the excitatory weights from the start, instead using a uniform coupling constant $\gamma$, which in our generic model case is the same as keeping $W=J$ ($g=1$) and varying both coupling strength at the same time. Uniform coupling $\gamma$, in some sense, hides the fact that inhibitory weight is not a control parameter of the AT transition for low connectivity networks. However, in \cite{Buendia2019}, the authors focus their attention on explaining the emergence of the Low-active intermediate (LAI) phase, which is an interesting behavior that emerges when inhibition is present. In our simulations, an LAI phase was also observed for both the Larremore and GGL models (in Fig.\ref{fig:2}.b, Fig.\ref{fig:3} and Fig.\ref{fig:4}.a, we described the LAI phase as the region where inhibition profusely modulates neuronal activity after AT takes place).

Studying the LAI phase in detail is outside the scope of our article. However, for completeness, we will also explore our results in the Larremore \textit{et al.} model. The simplified self-consistency equation for the Larremore \textit{et al.} model is,
\begin{equation}
    \rho = (1-\rho a)^{K_I}-\left(1-\eta\rho\right)^{K_E}\left(1-\rho b\right)^{K_I}
\end{equation}

As done in the GGL case, expanding to second order at $\rho=0$ yields
\begin{equation}
     \rho^{2}\left[\frac{K_E \eta^{2}(K_E-1)}{2} + K_I K_E \eta \right]+\rho (1- K_E \eta) = 0 \:,
\end{equation}
therefore,
\begin{equation}\label{eq:Larremore_AT}
    \rho^{*} \approx G(\eta,K_E)\left(K_E\eta-1\right)\:.
\end{equation}
which shows us that the AT transition occurs exactly as in the GGL model and is independent of inhibitory coupling strength. To test our analytical result, we simulate the Larremore \textit{et al.} model with linear and rational firing functions, for different $J$, $W$, $\Gamma$ and $K$, maintaining the relation between the excitatory and inhibitory population and the local connectivity in $8:2$. The results with the rational firing function are presented in Supplemental Materials II.A. 

\begin{figure*}[htb]
    \centering
    \includegraphics[width=.325\linewidth]{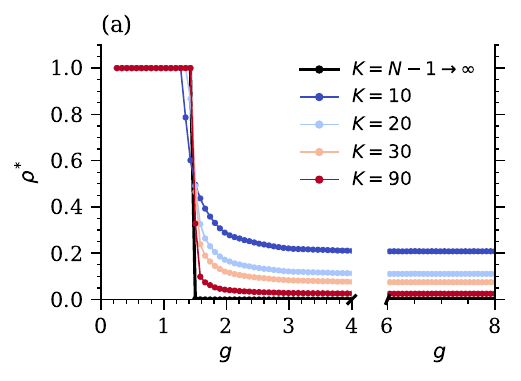}
    \includegraphics[width=.325\linewidth]{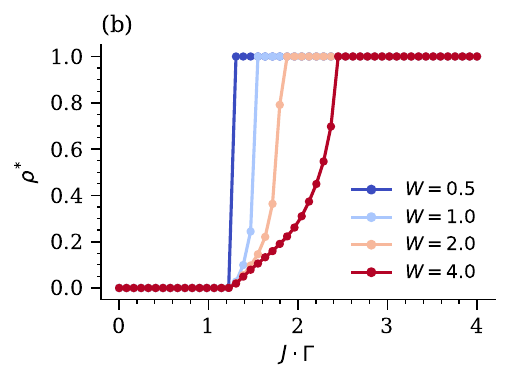}
    \includegraphics[width=.325\linewidth]{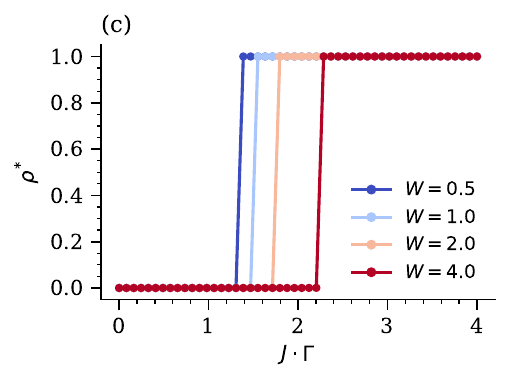}
    \caption{Larremore \textit{et al.} model with linear firing function. Steady-state activity $\rho^{*}$ as a function of (a) synaptic coupling ratio $g$ and (b,c) product between excitatory weights and gain. Simulations of the model in (a,b) random $K$-regular network and (a,c) complete graphs ($K=N-1$) with $N=10000$, $\Gamma=1$, $I=\theta=0$. (a) Complete graph MF does not correctly describe the behavior of sparse networks ($K\ll N$), activity becomes independent on $g$ for $g>>1$ when $K$ is small. (b) In random sparse networks, the intensity of activity is modulated by $W$ in the active phase, but the absorbing transition line does not depend on it. (c) In the complete graph, $W$ affects the critical point value, so it is a control parameter given a fixed value of $J$ and $\Gamma$.}
    \label{fig:5}
\end{figure*}

Our results are compatible with those obtained in \cite{Larremore2014}, where it was shown that the so-called branching function of the system is independent of the inhibitory contributions, besides some topology aspect (fraction of inhibitory neurons). Nonetheless, here we offer a complementary point of view of the phenomenon, by deriving the AT critical curve in terms of the different parameters of the model. We extend the results showing its validity in a more general model and address the impact of topological supposition over the absorbing phase control parameters and the AT transition. In Figure \ref{fig:5} we show that the Larremore \textit{et al.} model has the same behavior as the GGL. When the system is in the active phase, increasing $g$ is not capable of silencing the network activity unless the network has a complete graph topology (Fig.\ref{fig:5}.a). Modifying the inhibitory weight does not affect the critical point (Fig.\ref{fig:5}.b) unless the network has a complete graph topology (Fig.\ref{fig:5}.c). 

\begin{figure*}[htb]
    \centering
    \includegraphics{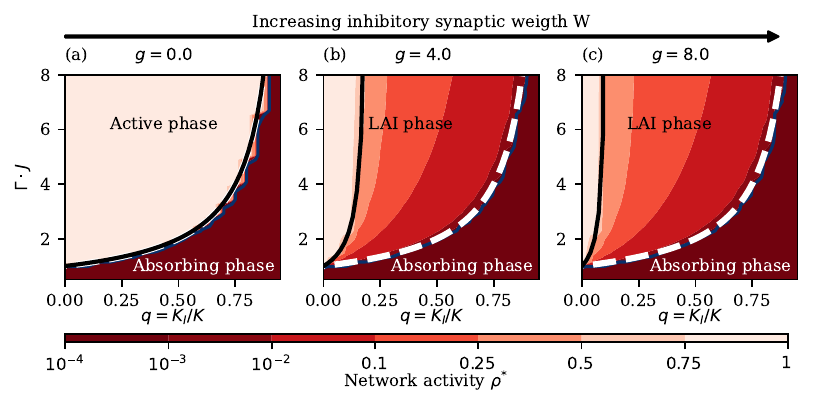}
    \caption{Phase diagram of Larremore \textit{et al.} model with Linear Firing function. Simulations in random $K$-regular network with $K=20$, $N=10000$, and different relative inhibitory weights $g=W/J$ with fixed $J=2$. The heat map shows the stationary activity of the network $\rho^{*}$, and the dashed lines are the critical curve obtained analytically by the tree-like MF approximation. The solid black line is the CGMF obtained for the GGL model Eq.\eqref{eq:Crict_GC}, which is also valid for the Larremore \textit{et al.} model. The absorbing transition is controlled by excitatory weight $J$, gain $\Gamma$, and the proportion of inhibitory neurons $q=K_I/K$, but not by the relative intensity of inhibitory weights $g=W/J$. The blue line is the contour of activity $\rho^{*}=5\times 10^{-4}$ obtained from the simulated data. It is possible to see the emergence of the LAI phase as shown in \cite{Buendia2019} when there is an inhibitory current, but the region of the stable absorbing state is not affected by it. The LAI phase is contained between the CGMF critical line and the tree-like MF, both curves were obtain analytically.}
    \label{fig:6}
\end{figure*}

Finally, the phase diagram of the Larremore (Fig.\ref{fig:6}) model obtained by simulations shows that the critical line of AT depends on the inhibitory proportion ($q$) and the product of gain and excitatory weights ($\Gamma\dot J$), but not on the inhibitory weight expressed here as the relative inhibitory weight or the E/I ratio $g$. However, as described by \cite{Buendia2019}, increasing inhibitory weights has an effect after the transition, causing an LAI phase between the absorbing phase and the fully ``active phase".

A similar phase diagram of Fig.\ref{fig:4} is presented for Shew and Lee \cite{Shew2020} with a modified Larremore \textit{et al.} model, where activity is measured by varying synaptic weight and relative inhibitory weight $g$. In that case, the authors use $g$ as a control parameter of a continuous transition, which seems to contradict our results. However, given a closer look at the Shew and Lee results, they are dealing with small networks ($N=1000$) with a high mean connectivity degree ($\avg{K}=200$), which leads its results closer to a complete graph than to a low connectivity sparse network. Nevertheless, analyzing Shew and Lee models using our results, we find that the authors are focusing not on the AT, but on what seems the remnant of LAI phase to fully active transition. The authors affirm that the Buendia \cite{Buendia2019} results on LAI phase do not apply in their case, because of the high connectivity, but using the insights learned from our results, we can see that the asynchronous irregular state studied by Shew and Li does seem to occur in the LAI phase or what is left of it. More details of this discussion are presented in Supplemental Material II.B.

After discussing our results in the GGL model and different versions of the Larremore \textit{et al.} model, we will now focus on the general case. For an arbitrary $\varphi(\{m_E,m_I\})$ function, the Eq.~\eqref{eq:selfcons_SN_sim_2} will have the same first term, but will also have a second term related to the transition $1 \rightarrow 0$. Thinking in the meaning of $\varphi(\{m_E,m_I\})$, it is plausible to propose that this probability will be independent of the neighbor states ($\{m_E,m_I\}$). So, in the simplest case, we will have a constant probability $\varphi$, and the simplified self-consistency equation will be,
\begin{equation}\label{eq:selfcons_SN_sim_2}
\rho = (1-\rho)\left[(1-\rho a)^{K_I}-\left(1-\eta\rho\right)^{K_E}\left(1-\rho b\right)^{K_I} \right] + \rho(1-\varphi)\:.
\end{equation}
Then, the expanded expression will be 
\begin{equation}\label{eq:selfcons_SN_sim_03}
\rho^{2}\left(K_E\eta +\frac{K_E\eta^{2}\left(K_E-1\right)}{2}+K_I K_E\eta\right)+\rho\left(K_E\eta - \varphi \right) = 0 \:.
\end{equation}

Solving Eq.~\eqref{eq:selfcons_SN_sim_03} we find the absorbing phase fixed point $\rho^{*}=0$ and, one more time
\begin{equation}
    \rho^{*} \approx G(\eta,\varphi,K_E)\left(\frac{K_E\eta}{\varphi}-1\right)\:,
\end{equation}
the critical curve one more time is independent of the inhibitory coupling strength, but now, it presents a dependence on the probability of inactivation after firing $\varphi$, which must be greater than 0. 

\begin{figure*}[htb]
\includegraphics[width=0.64
\linewidth]{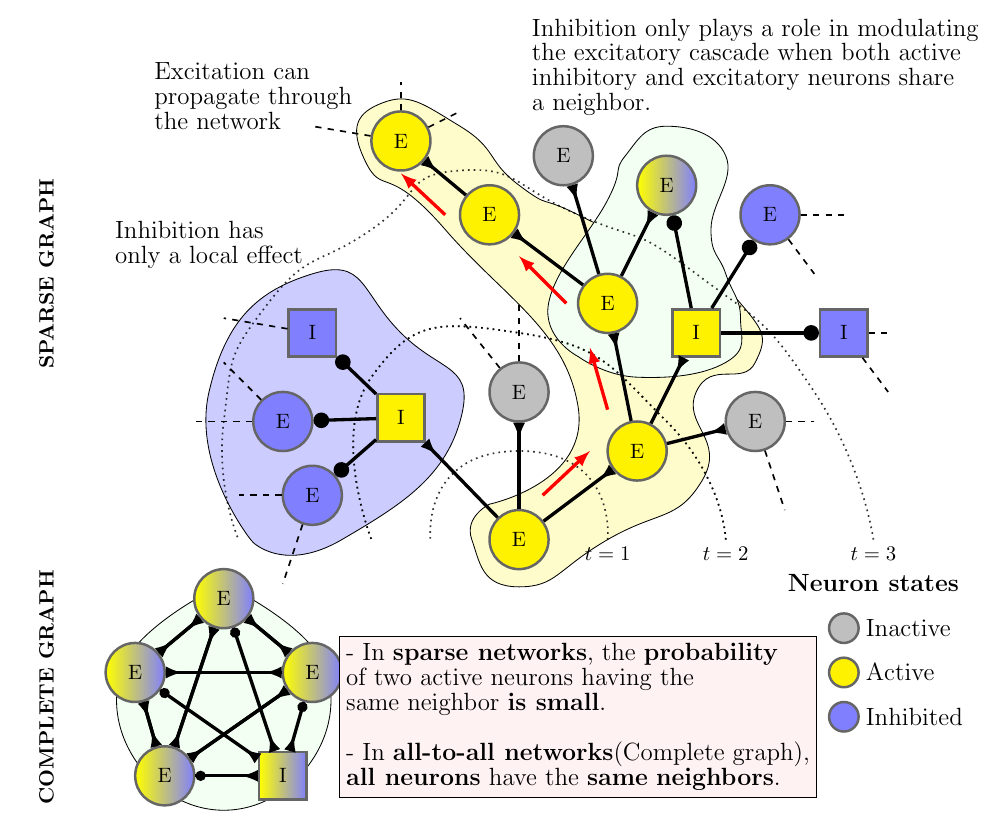}
\caption{\label{fig:7}Intuitive explanation of how topology influences system control parameters. Excitatory and inhibitory ``events" are fundamentally different: the first one could be propagated by the network by successive excitations, while the second one acts only locally, in first neighbors, and never propagates. However, when a complete graph topology is imposed, we artificially impose the same range of activities for both types of events; any excitatory or inhibitory event acts over all elements of the system, which will transform the inhibitory current into the control parameter of the absorbing phase transition, which will never happen in low connectivity networks.}
\end{figure*}

In conclusion, based on the analysis presented, we can affirm that our results are a general one, at least for this kind of stochastic neuron model, and it also seems to hold for Brunel's model when a careful look at the phase diagram is done. 

From a statistical physics point of view, there are some interesting details that we must be aware of: although both the sparse ($K>4$) and the complete graph networks have a dimensionality larger than the critical dimension $d_c$, and thus the critical exponents and scaling relations will be the same for the AT, the control parameters are not the same. Topological details affect the role the roles of parameters in the network, as intuitively explained in Fig.~\ref{fig:7}. The main reason is that excitatory events can be propagated through the network, generating neuronal cascades of activity, while inhibitory events have only a local effect, acting just over the nearest neighbors. Therefore, when all-to-all interactions are imposed, we artificially equate the effects of inhibitory and excitatory events by giving both a global range of action (each event has access to all neurons), which is a specific property of the CG topology. 

The explicit disappearance of the inhibitory/excitatory ratio $g$ in low-connectivity networks is our main
result. The behavior discussed here is a general one: In any spiking model, CG will artificially assign a greater role to inhibitory weights. The mean field derived from the CG assumption leads to erroneous conclusions about the system phase diagram. 

Finally, it is important to realize that the most connected neurons in biological systems have on the order of $10^{4}$ synapses but are immersed in networks with more than $10^{6}$ neurons, leading to a ratio $K/N \ll 1$. This fact makes low-connectivity sparse network models more relevant from the neuroscience point of view, adding relevance to the result presented here. Also, the independence of AT on inhibitory weights should be viewed as an important feature of sparse networks, as it will allow the system to self-tune properties such as average excitatory weights or neuronal gain to reach a near-critical regime, and, at the same time, have the freedom to adjust inhibitory weights to do computation or learning. As shown here, in sparse networks, less is different.

\begin{acknowledgments}
Gustavo Eduardo Mereles Menesse would like to thank the Programa Nacional de Becas de Postgrados en el Exterior ``Don Carlos Antonio López"(BECAL)-Paraguay for the financial support to his doctoral studies in the Physics and Mathematics Program of the University of Granada, and CAPES for financial support during the Master's studies.
O. K. acknowledges CNAIPS-USP and CNPq, Conselho Nacional de Desenvolvimento Científico e Tecnológico support.
\end{acknowledgments}

\bibliography{apssamp}
\bibliographystyle{apsrev4-1}

\end{document}


\title{Less is different: why sparse networks with inhibition differ from complete graphs}

\author{Gustavo Menesse}
\email{mereles930@gmail.com}
\thanks{Corresponding author}
\affiliation{Departamento de Electromagnetismo y Física de la Materia, Facultad de Ciencias, University of Granada, 18071 Granada, Spain}
\altaffiliation[Also at]{Departamento de Física, FFCLRP, Universidade de São Paulo, Ribeirão Preto, SP, 14040-901, Brazil }

\author{Osame Kinouchi}
\email{osame@ffclrp.usp}
\affiliation{
 Departamento de Física, FFCLRP, Universidade de São Paulo, Ribeirão Preto, SP, 14040-901, Brazil
}%

\date{\today}

\maketitle
\section{Analytical calculations}

\subsection{Complete graph mean field (Curie-Weiss mean field approximation)}

The following mean field follows the derivation presented in \cite{Girardi-Schappo2021}, with the only difference that here we use the rational firing function instead of linear using in the referenced paper. 

In the case of a complete graph in the thermodynamic limit, $K = N-1 \rightarrow \infty$, the equation \eqref{eq:Pot} of the main text becomes

\begin{eqnarray}\label{eq:Pot_MF}
    V^{E/I}_{i}[t + 1] &=& \Bigg[ \mu_i V^{E/I}_{i}[t] + I_i  + \left. \frac{1}{N}\left(
    \sum_{j=1}^{N_E}  J_{ij} X^{E}_j[t] -
    \sum_{j=1}^{N_I}  W_{ij} X^{I}_j[t] \right) \right]  \cdot \left[1 - {X_{i}}^{E/I}[t]\right].
\end{eqnarray}

As in the case of Bethe-Peierls approximation in the main text, assuming that our parameters are self-averaging, we define $J=\left< J_{ij}\right>$ and $W=\left< W_{ij}\right>$, $\Gamma=\left< \Gamma_{i}\right>$ and $\theta=\left< \theta_{i}\right>$. The activities of the populations are $\displaystyle \rho^{E}[t]=\frac{1}{N_E}\sum_{i}^{N_E} X_i^{E}[t]$ and $\displaystyle \rho^{I}[t]=\frac{1}{N_I}\sum_{i}^{N_I} X_i^{I}[t]$ and the fraction of excitatory and inhibitory neurons is $p=N_E/N$ and $q=1-p=N_I/N$, respectively. Therefore, taking the average of equation \ref{eq:Pot_MF} in the case of $\mu=0$, we have,
\begin{eqnarray}
V^{E/I}[t + 1] &=& \left[ I  + J p \rho^{E}[t]- W q \rho^{I}[t] \right] \cdot \left(1 - \rho^{E/I}[t]\right).
\end{eqnarray}
This system has two types of stationary states, one active, where $\rho^{E}=\rho^{I} \equiv \rho^{*} > 0$ and the silent state where $\rho^{E}=\rho^{I}\equiv \rho^{*}=0$. At any instant $t+1$ the active fraction is given by

\begin{equation}\label{eq:rho_MF}
    \rho^{E/I}[t+1]= \int_{\theta}^{\infty} \Phi(V)\mathbb{P}[t](V)dV \: ,
\end{equation}
where $\Phi(V)$ is the firing function and $\mathbb{P}[t](V)$ is the probability of having a neuron with membrane potential $V$ at time t. 

In the complete graph, the reset of the potential causes the $k$th sub-population of neurons that fires together to evolve together until a next fire, which allows us to write $\mathbb{P}[t](V)$ as
\begin{equation}
    \mathbb{P}[t](V) = \sum_{k=0}^{\infty}\eta^{E/I}_{k}\delta\left(V-U_{k}^{E/I}\right) \:,
\end{equation}
with $\delta(V)$ the Dirac's delta function, $U_{k}^{E/I}$ is the membrane potential of the $k$th population of excitatory or inhibitory neurons that fired $k$ time steps before $t$ and $\eta_{k}^{E/I}[t]$ is the proportion of such neurons with respect to the total excitatory or inhibitory population. This terms evolves in time as,
\begin{equation}
    \eta_{k+1}^{E/I}[t+1] = \left(1-\Phi(U_k^{E/I}[t]) \right)\eta_{k}^{E/I} \:,
\end{equation}
\begin{equation}
    U_{k+1}^{E/I}[t+1] = I[t] + pJ\rho^{E}[t]-qW\rho^{I}[t] \:.
\end{equation}
With this, the equation \eqref{eq:rho_MF} becomes,
\begin{equation}\label{eq:rho_MF_1}
\rho^{E/I}[t+1]= \sum_{k=0}^{\infty}\Phi(U_{k}^{E/I}[t])\eta_{k}^{E/I}[t]\: .
\end{equation}
If a neuron fires $k$ steps before time t, at time $t+1$ it can fire with probability $\Phi(U_{k}^{E/I})$ or become part of the population that fires $k+1$ time steps ago, which has density $\eta_{k+1}^{E/I}[t]$. When the stationary state is reached we have,
\begin{equation}\label{eq:rho_MF_2}
    \rho = \sum_{k=1}^{\infty}\eta_{k}\Phi(U_k) \:,
\end{equation}
\begin{equation}\label{eq:eta_0}
    \eta_{k} = \eta_{k-1}\left(1-\Phi(U_{k-1})\right) \:,
\end{equation}
\begin{equation}
    U_{k} = I + pJ\rho^{E}-qW\rho^{I} \:,
\end{equation}
remembering that $U_{0}=0$. From now on we will use \eqref{eq:eta_0} in \eqref{eq:rho_MF_2} and resetting the index we find ,
\begin{eqnarray}
    \rho &=& \sum_{k=0}^{\infty}\eta_{k}\left(1-\Phi(U_{k}) \right)\Phi(U_{k+1}) \nonumber \\ 
    &=& \Phi(U)\sum_{k=0}^{\infty}\eta_{k}\left(1-\Phi(U_{k}) \right)\nonumber \\ 
    &=& \Phi(U)\left(\sum_{k=0}^{\infty}\eta_{k}-\sum_{k=0}^{\infty}\eta_{k}\Phi(U_{k}) \right)\:.
\end{eqnarray}
Notice that, as we considered $\mu=0$ from the beginning, $U_{k+1}$ is not dependent on the history of the system, so we can discard the index $k+1$, we cannot do the same with $U_k$, because $k=0$ makes $U_0=0$. Using the fact that $\sum_{k=0}^{\infty}\eta_{k}=1$ and $\Phi(U_0)=0$, we obtain the self-consistency equation of the complete graph mean field,
\begin{equation}
\rho = \Phi(U)\left(1-\rho \right) \:.
\end{equation}

Substituting the rational firing function in the second factor,
\begin{equation}
\rho = \left(1-\rho \right)\frac{\Gamma(\bar{W}\rho+I-\theta)}{1+\Gamma(\bar{W}\rho+I-\theta)} \:,
\end{equation}
using $h=I-\theta$ and doing some algebra,
\begin{equation}
\rho = -2\Gamma\bar{W}\rho^{2}+(\Gamma\bar{W}-2\Gamma h) +\Gamma h =0\:,
\end{equation}

\subsection{Activity dynamics derivation in the Tree-like mean field approximation}

The probability of having an arbitrary neuron active in the time $t+1$ has two contributions, one from the jump $0 \rightarrow 1$ (inactive at time t to active at time $t+1$) and other from the probability of staying active $1 \rightarrow 1$, given all possible combinations of neighbor states $\{m_E,m_I\}$, which yields to
\begin{eqnarray}
    \mathbb{P}(X[t+1]=1) = \sum_{\{m_E,m_I\}}\mathbb{P}(X[t+1]=1|X[t]=0,\{m_E,m_I\})\mathbb{P}(X[t]=0,\{m_E,m_I\}) \nonumber \\ + \sum_{\{m_E,m_I\}}\mathbb{P}(X[t+1]=1|X[t]=1,\{m_E,m_I\})\mathbb{P}(X[t]=1,\{m_E,m_I\}) \: .
\end{eqnarray}

As neuron states $\{X[t]\}$ are independent of each other in the same instant of time $t$ (causal location), $\displaystyle\mathbb{P}(\{X[t]\})=\prod_{i} \mathbb{P}(X_i[t])$, so the joint probability $\mathbb{P}(X[t]=x,\{m_E,m_I\})=\mathbb{P}(X[t]=x)\cdot\mathbb{P}(\{m_E,m_I\})$, in other words, the state of a neuron at time $t$ is independent of the state of its neighbors at the same time $t$. With this, we have the following: 

\begin{small}
\begin{eqnarray} \label{Eq:ProbXact}
\mathbb{P}(X[t+1]=1) &=& \sum_{\{m_E,m_I\}}\left[\underbrace{\mathbb{P}(X[t+1]=1|X[t]=0,\{m_E,m_I\})}_{\Phi(\{m_E,m_I\})}\cdot \underbrace{\mathbb{P}(X[t]=0)}_{1-\rho[t]}\right. \nonumber \\ & & \left. +\underbrace{\mathbb{P}(X[t+1]=1|X[t]=1,\{m_E,m_I\})}_{1-\varphi(\{ m_E,m_I \})}\cdot \underbrace{\mathbb{P}(X[t]=1)}_{\rho[t]}\right] \cdot\mathbb{P}(\{m_E,m_I\}) \:.
\end{eqnarray}
\end{small}

A good estimation of the mean value is the empirical average of the state, which in the case of Boolean states is equal to the frequency of active neurons, and therefore equal to the network activity. Thus, at an arbitrary time $t$, we have $\mathbb{P}(X[t]=1)=\rho[t]$ and $\mathbb{P}(X[t]=0)=1-\rho[t]$. This said, from Eq.(\ref{Eq:ProbXact}) we obtain the activity
dynamics as:

\begin{eqnarray}\label{eq:rho_dyn}
\rho[t+1] &=& \sum_{\{m_E,m_I\}}\left[(1-\rho[t])\Phi(\{m_E,m_I\}) \right. +  \left. \rho[t]\left(1-\varphi(\{ m_E,m_I \})\right) \right]\cdot\mathbb{P}(\{m_E,m_I\}) \:.
\end{eqnarray}

\section{Larremore \textit{et al.} model}

\subsection{Larremore \textit{et al.} model with Rational Firing Function}

\begin{figure*}[htb]
    \centering
    \includegraphics[width=.325\linewidth]{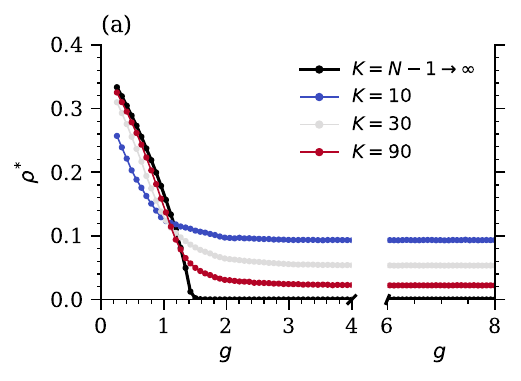}
    \includegraphics[width=.325\linewidth]{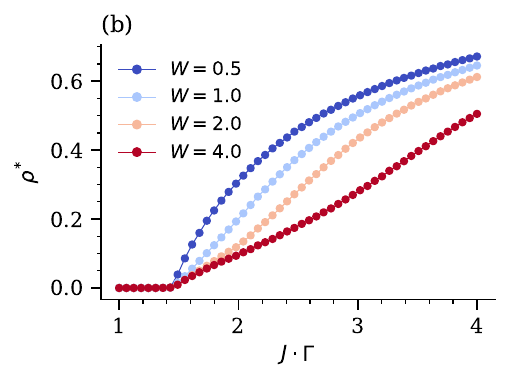}
    \includegraphics[width=.325\linewidth]{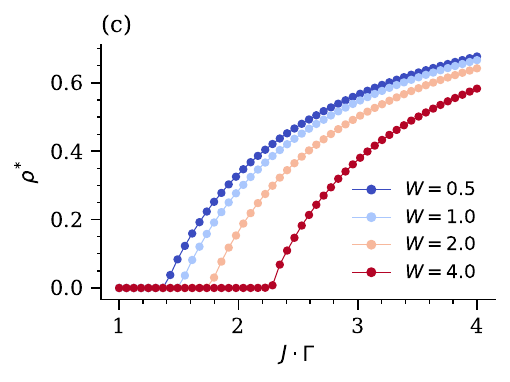}
    \caption{Steady state activity $\rho^{*}$ as a function of: (a) synaptic coupling ratio $g$ and (b,c) product between excitatory weights and gain. Simulations of the Larremore \textit{et al.} model with rational firing function in  (a,b) random $K$-regular network and (a,c) complete graphs ($K=N-1$) with $N=10000$, $\Gamma=1$, $I=\theta=0$ and $\mu=0$. Complete graph MF simulations (black line-dots). (a) Complete graph MF does not correctly describe the behavior of sparse networks ($K\ll N$), activity becomes independent on $g$ for $g>1$. (b) As seen for the GGL model, activity intensity is modulated by $W$ only after AT, but the AT critical point does not depend on $W$. (c) Similar to what was observed in GGL. In the complete graph, $W$ affects the value of the critical point, so it is a control parameter given a fixed value of $J$ and $\Gamma$. The LAI phase effect can be noticed in the sparse network. The presence of inhibition modify the activity curve, making it more linear with respect to the control parameter, a phenomenon not observed in the Complete Graph.}
    \label{fig:Sup1}
\end{figure*}

Simulations were performed with the rational firing function in the Larremore \textit{et al.} model. The critical line in this case is the same as in the GGL model when the rational firing function is used. But the discontinuous transition observed in pure excitatory networks disappears, given it place to a smooth transition similar to what is observed in GGL model. In Figure \ref{fig:Sup1} we observed the same behavior presented in the main text for GGL and Larremore with Linear Firing function. From Figure \ref{fig:Sup1}.a, sparse network activity becomes independent of $g$ for $g>>1$. In Figure \ref{fig:Sup1}.b, inhibition modulates activity only after the absorbing phase becomes unstable (after the AT occurs), generating a Low intermediate activity phase (LAI phase \cite{Buendia2019}). From Figure \ref{fig:Sup1}.c, inhibition becomes a control parameter of AT in the complete graph limit. 

\begin{figure*}[htb]
    \centering
    \includegraphics[width=\linewidth]{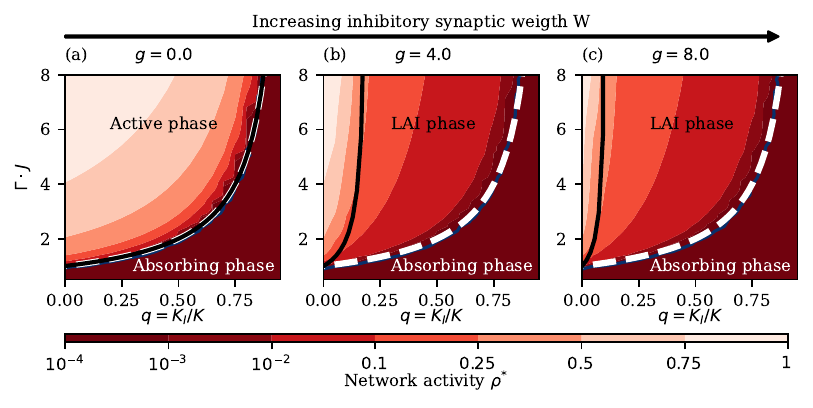}
    \caption{Phase diagram of Larremore \textit{et al.} model with Rational Firing function. Simulations in random $K$-regular network with $K=20$, $N=10000$ and different relative inhibitory weights $g=W/J$ with fixed $J=2$. The heat map shows the stationary activity of the network $\rho^{*}$, and the dashed lines are the critical curve obtained analytically by the tree-like MF approximation. The solid black line is the CGMF obtained for the GGL model Eq.\eqref{eq:Crict_GC}, which is also valid for the Larremore \textit{et al.} model. The absorbing transition is controlled by excitatory weight $J$, gain $\Gamma$, and the proportion of inhibitory neurons $q=K_I/K$, but not by the relative intensity of inhibitory weights $g=W/J$. The blue line is the contour of activity $\rho^{*}=5\times 10^{-4}$ obtained from the simulated data. Due to the rational function, there is no saturation and the system will never reach $\rho^{*}=1$. The delimitation of the LAI phase boundary is not as clean as with the linear function, but using the CGMF and tree-like MF critical curves, we can define the LAI phase boundaries.}
    \label{fig:Sup2}
\end{figure*}

The phase diagram of the Larremore \textit{et al.} model with the rational firing function (Fig.\ref{fig:Sup2}) shows that both CGMF and Tree-lik MF critical curves approximately agree when $W=g=0$ and $K>4$. In this case $K=20$, which leads us to a dimensionality higher than the critical one. However, only the Tree-like MF correctly captures the AT transition when inhibition is present. Between the CGMF and the Tree-like MF, we observe the LAI phase, region where the activity is modulated by the inhibition.  

\subsection{Shew and Li model}

Shew and Li \cite{Shew2020} present a modified Larremore \textit{et al.} model, with the difference that does not include a normalization factor for the inputs. The absence of normalization factor may be problematic, causing values as the nontrivial eigenvalue of the connectivity matrix and the eigenvalues radius explodes in the thermodynamic limits, as they present a dependence on the size of the system $N$. The absence of normalization factor will also cause the absorbent phase to disappear in the thermodynamic limit in a pure excitatory network, because the critical value of synaptic strength will tend to 0 in this limit. Nevertheless, it is possible to use our analytical derivation to obtain a critical curve for this model, also. In our general model, the normalization factor of synaptic inputs is K and the critical curve for the case of linear firing function is $(\Gamma J)_c = \frac{K}{K_E}$, Shew and Li use $\Gamma=1$, and since there is no normalization factor, the critical curve becomes $J_c = \frac{1}{K_E}$.

The authors use the same approach as Larremore to study the criticality, measuring the greater eigenvalue of connectivity matrix and observing when the eigenvalue becomes equal to 1. But this approach seems to have the limitation of not being able to distinguish the LAI phase transition to full active state (correctly described for the CGMF) from the absorbing phase transition (AT), correctly described by the tree-like MF in the low connectivity limit.

\begin{figure*}[htb]
    \centering
    \includegraphics[width=\linewidth]{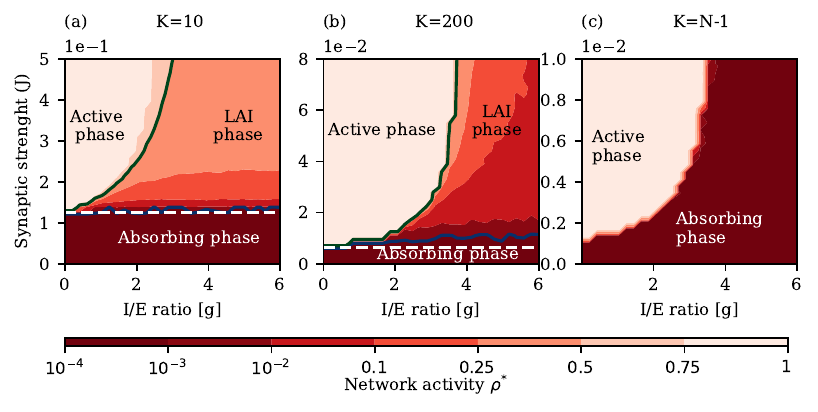}
    \caption{Phase diagram of the Shew and Li model in a increasingly high connected sparse network to complete graph: Stationary activity obtained in simulations on $T=10000$ time-steps on a K-regular random network (a,b) with $N=1000$ and inhibitory fraction $q=0.2$, where (b) same as used by Shew and Li \cite{Shew2020}, and (c) Complete Graph. White dashed line is the AT critical curve derived for the Shew and Lee model $J_c=\frac{1}{K_E}$, blue curve is the level curve of $\rho^{*}=1/N=10^{-3}$ and Green curve is the level curve $\rho^{*}=0.8$ presented as an estimate of the LAI phase - full active phase transition. Between the blue and green curves we observed what seem to be remnants of the LAI phase. The critical line for AT does not correctly describe the transition as the system has a high connectivity and the inhibitory current is starting to control the AT transition as can be seen in the inclination of the blue line in (b). (c) In complete graph, the LAI phase disappears becoming part of the absorbing phase, and the AT transition is described by the CGMF.}
    \label{fig:Sup3}
\end{figure*}

Our derived critical curve correctly captures the region where the absorbing phase loses stability, as shown in Fig.\ref{fig:Sup3}.a. However, as we increase connectivity K with respect to network size N, the system slowly begins to approach the CGMF results. As shown by Buendia \cite{Buendia2019}, for a high enough connectivity, the LAI phase should disappear and the LAI phase to a full active phase transition will become the AT, which in these limits will be controlled by the g. Shew and Li \cite{Shew2020} discuss the Buendia article explaining that the LAI phase only appears for $K/N<0.01$, in fact, in the Buendia's article, the authors claim that the LAI phase exists until $K/N \ge 0.03$, moment from which their interpolated critical curve coincides with their analytic critical curve. However, as we shown here, it seems that this modified Larremore \textit{et al.} model has an extended region of the LAI phase, so the Asynchronous Irregular activity described by Shew and Lee could be a property of the LAI phase. A more detailed study is required to address these questions.

\bibliography{apssamp}
\bibliographystyle{apsrev4-1}